\documentclass[%
 amsmath,amssymb,
 aps,
 prd,
twocolumn, superscriptaddress
]{revtex4-1}

\usepackage{float}
\usepackage[normalem]{ulem}
\usepackage[dvipsnames]{xcolor}

\usepackage{graphicx}
\usepackage{dcolumn}
\usepackage{bm}
\usepackage{hyperref}

\usepackage{xcolor}
\usepackage{orcidlink}
\hypersetup{colorlinks, citecolor=BrickRed}
\hypersetup{colorlinks=true, linkcolor=Blue, urlcolor=MidnightBlue}

\begin{document}

\preprint{APS/123-QED} 

\title{Compact stars in Rastall gravity: hydrostatic equilibrium and radial pulsations }

\author{Juan M. Z. Pretel \orcidlink{0000-0003-0883-3851}}
 \email{juanzarate@cbpf.br}
 \affiliation{Centro Brasileiro de Pesquisas F{\'i}sicas, Rua Dr.~Xavier Sigaud, 150 URCA, Rio de Janeiro CEP 22290-180, RJ, Brazil
}
\affiliation{
 Grupo de investigaci{\'o}n ``Gravitaci{\'o}n, Cosmolog{\'i}a, Campos y Cuerdas'', 
 Universidad Nacional de Trujillo, Departamento de F{\'i}sica, Av.~Juan Pablo II S/N; Ciudad Universitaria, Trujillo, La Libertad, Peru
}

\author{Cl\'esio E. Mota \orcidlink{0000-0002-8616-0894}}
 \email{clesio200915@hotmail.com}
 \affiliation{
 Departamento de F{\'i}sica, CFM - Universidade Federal de Santa Catarina, C.P. 476, Florianópolis, CEP 88.040-900, SC, Brazil
}

\date{\today}

\begin{abstract}
Within the context of Rastall gravity, we investigate the hydrostatic equilibrium and dynamical stability against radial pulsations of compact stars, where a free parameter $\beta$ measures the deviations from General Relativity (GR). We derive both the modified Tolman-Oppenheimer-Volkoff (TOV) equations and the Sturm-Liouville differential equation governing the adiabatic radial oscillations. Such equations are solved numerically in order to obtain the compact-star properties for two realistic equations of state (EoSs). For hadronic matter, the fundamental mode frequency $\omega_0$ becomes unstable almost at the critical central energy density corresponding to the maximum gravitational mass. However, for quark matter, where larger values of $\vert\beta\vert$ are required to observe appreciable changes in the mass-radius diagram, there exist stable stars after the maximum-mass configuration for negative values of $\beta$. Using an independent analysis, our results reveal that the emergence of a cusp can be used as a criterion to indicate the onset of instability when the binding energy is plotted as a function of the proper mass. Specifically, we find that the central-density value where the binding energy is a minimum corresponds precisely to $\omega_0^2 =0$.

\end{abstract}

\maketitle


\section{Introduction}

Although it is very common to consider the conservation of energy–momentum tensor in gravity theories, it has been argued that such conservation could be violated in quantum systems \cite{Minazzoli2013, Harko2014}. See e.g.~Ref.~\cite{Velten2021} for a review about alternative gravity theories in which the null covariant divergence of the energy–momentum tensor is not achieved. In particular, by assuming that the covariant divergence of the energy-momentum tensor is no longer zero, but proportional to the covariant derivative of the Ricci scalar, Rastall modified the GR theory fifty years ago \cite{Rastall1972}. This assumption describes the non-minimal coupling between matter and geometry, and its cosmological consequences have been previously studied in the literature \cite{Rawaf1996, Batista2012, Batista2013, Moradpour2016, Moradpour2017PRD, Khyllep2019, Singh2020, Akarsu2020, Tang2020, Singh2022}. Manna and collaborators \cite{Manna2020} investigated the classical tests (such as precession of perihelion, deflection of light and time delay) in Rastall theory. Furthermore, at cosmological level, we must emphasize that generalized versions for original Rastall gravity have also been explored \cite{Moradpour2017, Das2018, Lin2020, Shahidi2021, Shabani2022}. The propagation of axial gravitational waves in the spatially flat conformal FLRW spacetime within the framework of generalized Rastall theory was discussed by Salti \cite{Salti2020}.

On the other hand, in the strong-field regime, various authors have reported black hole solutions in Rastall gravity \cite{Heydarzade2017CJP, Kumar2018, Lin2019, Kumar2021}, as well as their thermodynamic properties \cite{Lobo2018}. It was shown that Kiselev-like black holes surrounded by regular matter in Rastall theory may be considered as Kiselev black holes surrounded by exotic matter in GR \cite{Heydarzade2017}. The quasinormal oscillations corresponding to the black hole in presence of non-linear electrodynamic sources revealed significant deviations from a general charged black hole in Rastall gravity under certain conditions \cite{Gogoi2021}. More recently, Shao \textit{et al.}~\cite{Shao2022} investigated the quasinormal modes and late-time tail of massless scalar perturbations of a magnetized black hole, while the Hawking radiation of a Schwarzschild black hole surrounded by a cloud of strings in Rastall theory was discussed in Ref.~\cite{Li2022}. Moreover, by means of Newman-Janis algorithm without complexification, an exact solution of Kerr black hole surrounded by a cloud of strings in Rastall gravity was obtained in Ref.~\cite{Li2021PRD}.

Over the last years, there has been a growing interest in constructing compact stars within the context of Rastall gravity. As a matter of fact, the equilibrium structure of neutron stars described by an isotropic perfect fluid was examined in Refs.~\cite{Oliveira2015, Xi2020, Mota2022IJMPD}, and the tidal Love numbers of such stars were calculated in \cite{Meng2021}. Anisotropic compact stars under the same gravitational framework were also investigated in the past \cite{Abbas2018, Mota2022, Nashed2022, Tangphati2022}, as well as taking into account electric charge \cite{Shahzad2019, Salako2022GRG, Mustafa2022CJP} and rapid rotation effects \cite{Silva2021}. In addition, the gravitational collapse process of a homogeneous perfect fluid and the dynamics of dissipative collapse of a self-gravitating source (using the M{\"u}ller–Israel–Stewart approach) were explored in Refs.~\cite{Ziaie2019} and \cite{Tahir2021}, respectively. Exotic objects as wormholes in Rastall theory were also studied in \cite{Halder2019, Mustafa2020}.

Theoretically, the radial stability of a compact star can be investigated by considering small perturbations from the hydrostatic equilibrium. In Einstein gravity, Chandrasekhar was the first to derive the linearized equations describing the adiabatic radial oscillations of isotropic compact stars \cite{ChandrasekharApj, Chandrasekhar1}. Specifically, the system of differential equations is treated as an eigenvalue problem and, subsequently, the emergence of real (complex) discrete frequencies is attributed to a radial stability (instability) of the star. For further reading on radial oscillations of compact stars in GR we also refer the reader to Refs.~\cite{Chanmugam1977, Benvenuto1991, Gondek1997, KokkotasR2001, Flores2010, Pereira2018, Pretel2020MNRAS, Clemente2020, Bora2021, Sun2021, Jimenez2021, Hong2022}. Additionally, this subject has also been addressed in some modified theories of gravity, see e.g.~Refs.~\cite{Sham2012, Dzhunushaliev2020, Pretel2021JCAP, Sarmah2022}. To the best of our knowledge, the dynamical stability of compact stars (adopting a perturbative approach to the field equations) has not yet been investigated in Rastall gravity, so the purpose of the present work is to fill this gap. To do so, we will first derive the modified TOV equations and they will be solved for two EoSs describing quark matter and hadronic matter. Our second task will be to perturb all equations involved up to first order in the metric and fluid variables in order to obtain the radial pulsation equations.

Before continuing with our study, it is important to comment on some criticisms about Rastall gravity, which have been raised in the paper \cite{Visser2018}. Visser claimed that Rastall gravity is a trivial rearrangement of the matter sector in Einstein gravity; as gravity there is absolutely nothing
new. This argument is not true for one simple reason: In any modified gravity theory it is possible to rearrange the field equations and rewrite them in a traditional form as the Einstein equations. For example, the field equations in the well-known $f(R)$ gravity theories \cite{SotiriouFaraoni,Felice}
\begin{equation}
f_R R_{\mu\nu} - \dfrac{1}{2}g_{\mu\nu}f - \nabla_\mu\nabla_\nu f_R + g_{\mu\nu}\square f_R = 8\pi T_{\mu\nu}^{\rm ord} 
\end{equation}
can be written as $G_{\mu\nu}= 8\pi T_{\mu\nu}^{\rm tot}$, where the total energy-momentum tensor $T_{\mu\nu}^{\rm tot} = T_{\mu\nu}^{\rm ord}+ T^{\rm cur}_{\mu\nu}$ is composed of two fluids indeed \cite{Capozziello2012, Pretel2022JCAP}: a standard matter fluid (i.e., ordinary matter described by $T_{\mu\nu}^{\rm ord}$) and a curvature fluid (also called in the literature as effective fluid), being the latter described by \cite{Pretel2022JCAP}
\begin{align}
    T^{\rm cur}_{\mu\nu} =&\ \dfrac{1}{8\pi}\left[ (1 - f_R)R_{\mu\nu} + \dfrac{1}{2}(f -R)g_{\mu\nu}  \right.  \nonumber \\
    &\hspace{0.8cm} + \nabla_\mu\nabla_\nu f_R - g_{\mu\nu}\square f_R \bigg] .
\end{align}

One can observe that equation $G_{\mu\nu}= 8\pi T_{\mu\nu}^{\rm tot}$ does not mean that GR is equivalent to $f(R)$ gravity theory. In fact, such equivalence only exists when $f(R)= R$. If we followed Visser's argument, we would have that the $f(R)$ theories contribute absolutely nothing new by simply rearranging the field equations, however, as we already know, that is false! Therefore, Rastall gravity is not equivalent to Einstein gravity. It is important to mention that in addition to this argument, there are other issues showing that these two theories are not equivalent. As a matter of fact, Darabi and collaborators \cite{Darabi2018} have obtained different conclusions from Visser's ones, and they have indeed argued that Rastall gravity is an ``open'' theory when compared to GR. Specifically, the authors have stated that the ordinary definition of energy-momentum tensor is valid in the Rastall theory, unlike the claims of Visser \cite{Visser2018}.

Within the gravitational framework of GR, the $M(\rho_c)$ method is often compatible with the analysis of the normal vibration modes, namely, the maximum-mass value on the $M(\rho_c)$ curve and the zero fundamental mode frequency are found at the same central energy density. Nevertheless, these methods are not compatible in more realistic or general situations; such as in anisotropic stars \cite{Pretel2020EPJC} unless the tangential pressure is held fixed at the surface of the star \cite{Arbanil2016}, color superconducting quark stars \cite{Flores2010}, hybrid stars composed by a quark matter core and hadronic external layers \cite{Pereira2018}, neutron stars with a dark-energy core \cite{Pretel2024PRD}, charged strange quark stars \cite{Arbanil2015}, and polytropic configurations in modified gravity \cite{Pretel2021JCAP}. As we will see later, the gravitational mass of a star in Rastall gravity does not only depend on the energy density, but there is also an extra mass contribution due to the Rastall term, also treated as ``effective fluid'' in modified gravity. In that regard, the standard $M(\rho_c)$ method is expected to be invalid in the Rastall scenario. An indicator that could also establish the onset of instability is the formation of a cusp when the gravitational binding energy (as a function of the baryon rest mass) is a minimum. As a matter of fact, this analysis was carried out in scalar-tensor theories of gravity \cite{Doneva2013, PRETEL2024PDU}, 4D Einstein-Gauss-Bonnet gravity \cite{Doneva2021}, $f(R,T)$ theories with conserved energy-momentum tensor \cite{Pretel2021JCAP}, and within the context of $R$-squared gravity \cite{Jimenez2022}. In this study, we will therefore consider independent approaches (i.e., by analyzing the binding energy and calculating the fundamental vibration mode frequencies) in order to obtain rigorous results on the radial stability of compact stars in Rastall gravity.

The plan of this work is as follows: In Sec.~\ref{Sect2} we briefly review Rastall gravity, we explicitly express the field equations for a spherically symmetric system, derive the corresponding modified TOV equations, and we present the equations of state. Considering small adiabatic perturbations from the equilibrium state, in Sec.~\ref{Sect3} we address the dynamical stability and we derive the radial oscillation equations. Sec.~\ref{Sect4} presents a discussion of the numerical results for the equilibrium configurations as well as an analysis about the frequencies of normal vibration modes. Finally, our conclusions are presented in Sec.~\ref{Sect5}. We will use a geometric unit system and the sign convention $(-,+,+,+)$ throughout the text. However, our results will be given in physical units for comparison reasons.


\section{Stellar structure equations}\label{Sect2}

We suppose that spacetime geometry of a spherically symmetric stellar configuration is described by the well-know line element
\begin{equation}\label{metric}
ds^2 = g_{\mu\nu}dx^\mu dx^\nu = -e^{2\psi}dt^2 + e^{2\lambda} dr^2 + r^2d\Omega^2, 
\end{equation}
where $g_{\mu\nu}$ is the metric tensor, $x^\mu = (t, r, \theta, \phi)$ are the Schwarzschild-like coordinates, and the metric functions $\psi$ and $\lambda$, in principle, depend on the coordinates $t$ and $r$. Moreover, $d\Omega^2 = d\theta^2 + \sin^2\theta d\phi^2$ is the line element on the unit 2-sphere.

With respect to the matter-energy distribution, we assume that the compact star is made of an isotropic perfect fluid, namely
\begin{equation}\label{EMtensor}
    T_{\mu\nu} = (\rho + p)u_\mu u_\nu + pg_{\mu\nu} ,
\end{equation}
with $\rho$ being the energy density, $p$ the pressure, and $u_\mu$ is the four-velocity of the fluid, given by 
\begin{equation}\label{4Velocity}
    u^\mu = \frac{dx^\mu}{d\tau} = u^0\left( 1, \frac{dx^i}{dt} \right) ,
\end{equation}
where $\tau$ is the proper time and $u_\mu u^\mu = -1$. Following the same procedure carried out by Chandrasekhar \cite{ChandrasekharApj, Chandrasekhar1} in GR theory, we consider a spherically symmetric system with motions, if any, only in the radial directions, so that $u^\mu= (u^0, u^1, 0, 0)$ and $T_2^{\ 2} = T_3^{\ 3} =p$.

Under the assumption that the energy-momentum tensor is not conserved in curved spacetime, Rastall proposed a generalization of Einstein's conventional theory \cite{Rastall1972}. Indeed, within the framework of Rastall gravity, the modified field equations are given by
\begin{equation}\label{FieldEq0}
    R_{\mu\nu} - \frac{\alpha}{2}g_{\mu\nu}R = 8\pi T_{\mu\nu} ,
\end{equation}
or alternatively
\begin{equation}\label{FieldEq1}
    G_{\mu\nu} = 8\pi \left[ T_{\mu\nu}- \frac{1-\alpha}{16\pi}Rg_{\mu\nu} \right] ,
\end{equation}
where $R_{\mu\nu}$ is the Ricci tensor, $R$ denotes the scalar curvature, $G_{\mu\nu}$ is the Einstein tensor and $\alpha$ is the so-called Rastall parameter. Notice that the curvature itself contribute to the total energy of the system, as postulated by Rastall. As is evident, the standard GR theory is retrieved when $\alpha =1$.

By taking the trace of Eq.~(\ref{FieldEq1}), we get $R = 8\pi T/(1-2\alpha)$ and hence the field equations can be written as 
\begin{equation}\label{FieldEq2}
    G_{\mu\nu} = 8\pi T_{\mu\nu} - 8\pi\beta Tg_{\mu\nu} ,
\end{equation}
where we have defined $\beta \equiv \frac{1-\alpha}{2(1 - 2\alpha)}$.

Maintaining the validity of the gravitational Bianchi identity (i.e., $\nabla_\mu G^{\mu\nu} =0$), the four-divergence of Eq.~(\ref{FieldEq2}) leads to 
\begin{equation}\label{4diver}
    \nabla_\mu T^{\mu\nu} = \beta\nabla^\nu T = \frac{1- \alpha}{16\pi} \nabla^\nu R .
\end{equation}
Rastall questioned the traditional assumption that the covariant derivative of the energy-momentum tensor is zero in curved spacetime \cite{Rastall1972}. As we can see in Eq.~(\ref{4diver}), he considered that the four-divergence of the energy-momentum tensor is proportional to the variation of the Ricci scalar. It is therefore evident that the energy-momentum tensor is a non-conserved quantity. Nevertheless, the assumption $\nabla_\mu T^{\mu\nu} \neq 0$ opens the possibility of a gravitationally induced particle production as shown by Harko and collaborators \cite{Harko2014, Harko2015EPJC}. In fact, there exists a considerable number of modified gravity theories allowing departures from the usual conservative context. For a review of non-conservative gravity theories, we refer the reader to Ref.~\cite{Velten2021}.

For the index $\nu= 1$, the non-conservative equation $\nabla_\mu T_1^{\ \mu} = \beta\nabla_1 T$ becomes
\begin{align}\label{4diverA}
\partial_t T_1^{\ 0} + \partial_r T_1^{\ 1} &+ T_1^{\ 0}(\dot{\psi} + \dot{\lambda}) + \left[T_1^{\ 1} - T_0^{\ 0}\right]\psi'  \nonumber  \\
&+ \dfrac{2}{r}\left[T_1^{\ 1} - p\right] = \beta \partial_r (-\rho + 3p) , 
\end{align}
where overdots and primes denote partial differentiation with respect to $t$ and $r$, respectively. If $\beta= 0$, the conservation law of energy and momentum is restored and Rastall gravity falls back to Einstein's theory.

Furthermore, the non-null components of the field equations (\ref{FieldEq2}) are given by 
\begin{align}
 &\dfrac{1}{r^2}\partial_r(re^{-2\lambda}) - \dfrac{1}{r^2} = 8\pi T_0^{\ 0} + 8\pi\beta (\rho- 3p) ,   \label{FEq1}   \\  
 &e^{-2\lambda}\left( \dfrac{2}{r}\psi' + \dfrac{1}{r^2} \right) - \dfrac{1}{r^2} = 8\pi T_1^{\ 1} + 8\pi\beta (\rho -3p) ,   \label{FEq2}    \\
 &e^{-2\lambda}\left[ \psi'' + \psi'^2 - \psi'\lambda' + \dfrac{1}{r}(\psi' - \lambda') \right]  \nonumber  \\
 &\quad + e^{-2\psi}\left[ \dot{\lambda}\dot{\psi} - \ddot{\lambda} - \dot{\lambda}^2 \right] = 8\pi p + 8\pi\beta (\rho -3p) ,  \label{FEq3}    \\
 &\dfrac{2}{r}e^{-2\lambda}\dot{\lambda} = 8\pi T_0^{\ 1} ,  \label{FEq4}
\end{align}
and, by means of Eqs.~(\ref{FEq1}) and (\ref{FEq2}), we can obtain the following relation which will be used later: 
\begin{equation}\label{UsefulEq}
\frac{2}{r}e^{-2\lambda}(\psi'+\lambda') = 8\pi(T_1^{\ 1} - T_0^{\ 0}) .
\end{equation}

\subsection{Modified TOV equations}

When the stellar fluid remains in a state of hydrostatic equilibrium, none of the variables depend on the time coordinate.
This implies that $u^\mu = (u^0, 0, 0, 0)$, $T_0^{\ 0} = -\rho_0$ and $T_1^{\ 1} = T_2^{\ 2} = T_3^{\ 3} = p_0$, where the lower index $0$ indicates that the quantities are evaluated in the equilibrium. As a consequence, Eqs.~(\ref{FEq1}), (\ref{FEq2}), (\ref{4diverA}) and (\ref{UsefulEq}) assume the form, respectively, 
\begin{eqnarray}
  \frac{d}{dr}(re^{-2\lambda_0}) &=& 1- 8\pi r^2\rho_0 + 8\pi\beta r^2(\rho_0- 3p_0) , \label{ModEq1}  \\
  \frac{2}{r}e^{-2\lambda_0}\psi'_0 &=&\frac{1 - e^{-2\lambda_0}}{r^2} + 8\pi p_0 + 8\pi\beta(\rho_0 - 3p_0) , \qquad  \label{ModEq2}  \\
  \frac{dp_0}{dr} &=& -(\rho_0+ p_0)\psi'_0 - \beta(\rho'_0 - 3p'_0) , \label{ModEq3}  \\
  \frac{2}{r}(\psi'_0 + \lambda'_0) &=& 8\pi(\rho_0 + p_0)e^{2\lambda_0} . \label{ModEq4}
\end{eqnarray}

Integrating Eq.~(\ref{ModEq1}) we obtain the usual expression
\begin{equation}\label{Massfun1}
    e^{-2\lambda_0} = 1 - \frac{2m}{r} ,
\end{equation}
where $m(r)$ stands for the gravitational mass within a sphere of radius $r$, given by 
\begin{equation}\label{Massfun2}
    m(r) = 4\pi\int_0^r \bar{r}^2\rho_0(\bar{r})d\bar{r} - 4\pi\beta\int_0^r \bar{r}^2 \left[ \rho_0(\bar{r}) - 3p_0(\bar{r}) \right]d\bar{r} ,
\end{equation}
or alternatively, $m= m_\rho+ m_{\rm eff}$, where $m_\rho$ is the standard mass widely known in GR and $m_{\rm eff}$ is an extra effective mass due to the modification of the field equations. In other words, the Rastall term contributes additional mass, and as we will see later in our results, it ends up modifying the radial stability of a compact star. As expected, when $\beta =0$ we recover the typical expression obtained in Einstein gravity. At the surface, where the pressure vanishes, $m(r_{\rm sur}) \equiv M$ is the total mass of the star. Taking into account Eq.~(\ref{Massfun1}), from Eq.~(\ref{ModEq2}) we obtain
\begin{equation}\label{DervPsiEq}
 \frac{d\psi_0}{dr} = \left[ \frac{m}{r^2} + 4\pi rp_0 + 4\pi\beta r(\rho_0 - 3p_0) \right] \left( 1- \frac{2m}{r} \right)^{-1} .
\end{equation}

Thus, in view of Eqs.~(\ref{ModEq3}), (\ref{Massfun2}) and (\ref{DervPsiEq}), the relativistic structure of a compact star within the framework of Rastall gravity is described by the modified TOV equations:
\begin{align}
    \frac{dm}{dr} =&\ 4\pi r^2\rho - 4\pi\beta r^2(\rho- 3p) ,  \label{TOV1}  \\
    \frac{dp}{dr} =& -\frac{\rho + p}{1- 3\beta} \left[ \frac{m}{r^2}+ 4\pi rp + 4\pi\beta r(\rho- 3p) \right]  \nonumber  \\
    &\times \left( 1- \frac{2m}{r} \right)^{-1} - \frac{\beta}{1- 3\beta}\frac{d\rho}{dr}  ,  \label{TOV2}  \\
    \frac{d\psi}{dr} =& -\frac{1- 3\beta}{\rho+ p}\frac{dp}{dr} - \frac{\beta}{\rho+ p}\frac{d\rho}{dr}  \label{TOV3} ,
\end{align}
where we have removed the subscript zero because all quantities correspond to the state of hydrostatic equilibrium. This set of differential equations plays a crucial role in describing the hydrostatic equilibrium of compact stars in Rastall theory. They are also known as stellar structure equations and, as in any gravity theory, they allow us to obtain the most basic macroscopic properties of a star such as its radius and gravitational mass. Similar to the GR context, they are the base equations for further analysis of moment of inertia, oscillation spectrum and tidal properties. Recent studies on the relativistic structure of compact objects in modified gravity can be found in Refs.~\cite{Xi2020, Mota2022IJMPD, Meng2021, Abbas2018, Mota2022, Nashed2022, Tangphati2022, Shahzad2019, Salako2022GRG, Mustafa2022CJP, Silva2021, Pretel2022JCAP, PRETEL2024PDU, Jimenez2022, Shamir2022, Pretel2022MPL, Bhattacharya2023, Karmakar2023, Oikonomou2023CQG, Oikonomou2023MNRAS, Numajiri2023, Nashed2023EPJC, Rashid2023, Rashid2023FP, Bhar2023PDU, Bhar2023FP, Tangphati2023, Gammon2024, Malik2024, Sedaghat2024}. This work aims to carry out a rigorous analysis about the radial stability of compact stars in Rastall gravity through radial and adiabatic pulsations. In other words, the static background, described by the modified TOV equations (\ref{TOV1})-(\ref{TOV3}), will be subjected to small deviations from equilibrium. We will return to this in section \ref{Sect3}.

As expected, when the free parameter $\beta$ vanishes (this is, $\alpha =1$), one retrieves the conventional TOV equations in the pure GR case \cite{Tolman1939, Oppenheimer1939}. The variables to be determined by this system of equations are $m(r)$, $\rho(r)$, $p(r)$ and $\psi(r)$, while the metric function $\lambda(r)$ can be determined from Eq.~(\ref{Massfun1}). Given an EoS in the form $p= p(\rho)$, Eqs.~(\ref{TOV1}) and (\ref{TOV2}) can be integrated for a specific value of central energy density and by guaranteeing regularity at the center of the star. Besides, since $R=0$ outside the star, we can still use the Schwarzschild vacuum solution to describe the exterior spacetime. Namely, this allows the interior solution to be matched to the exterior Schwarzschild solution at the boundary $r= r_{\rm sur}$. Therefore, the system of differential equations (\ref{TOV1})-(\ref{TOV3}) can be solved by imposing the following boundary conditions
\begin{align}\label{BConditions}
\rho(0) &= \rho_c,   &   m(0) &= 0,   &   \psi(r_{\rm sur}) &= \frac{1}{2}\ln\left[ 1 - \frac{2M}{r_{\rm sur}} \right] . \ \
\end{align}

The most basic macroscopic properties of a compact star such as mass and radius will be calculated for a range of central-density values and by considering different values of the Rastall parameter. Other observables such as the gravitational redshift of light emitted at the surface of the equilibrium star can also be analyzed in terms of the parameter $\beta$. This quantity is given by
\begin{equation}\label{RedShiftEq}
    z_{\rm sur} = \left[ 1- \frac{2M}{r_{\rm sur}} \right]^{-1/2} -1 .
\end{equation}

\subsection{Equations of state}\label{sec_EoS}

In this subsection, we discuss the EoSs used in the present work to describe quark matter and hadronic matter. The first is then used to investigate quark stars and the second, hadronic stars. A didactic and more extensive explanation of the EoS can be found in Ref.~\cite{Menezes2021} and references therein. These EoSs will be the input to the modified hydrostatic equilibrium and radial oscillation equations in Rastall gravity.

\subsubsection{Quark matter}\label{sub_quarks}

To investigate quark stars we use a simple relativistic model to describe quark matter, the MIT bag model \cite{Chodos1974}. In general terms, such model describes the confinement of quarks in a volume of space capable of containing hadronic fields. Inside the bag, creating and maintaining this region in a vacuum requires constant positive potential energy per unit volume, namely {\it the bag constant} ($B$). Inside this volume, the moving quarks have an associated kinetic energy and no colour currents survive on the surface. We therefore assume the quarks in the interior of the bag as a Fermi gas whose energy at the border of the bag is negligible when compared with the energies inside it. 

The Lagrangian density that reproduces the dynamics of the quarks $\psi_{q}$ contained in a bag of volume $V$ delimited by the surface $S$ can be written as
\begin{equation}\label{lagrangian_density}
\mathcal{L}_{\rm MIT} = \sum_{q} \left[\bar{\psi}_{q}(i\gamma^{\mu}\partial_{\mu} - m_{q})\psi_{q} - B\right]\Theta_{V} - \frac{1}{2}\bar{\psi}_{q}\psi_{q}\delta_{S},
\end{equation}
where $q$ denotes the flavours of the quarks involved, of masses $m_{q}$; $B$ is the bag constant; $\Theta_{V}$ is the Heaviside function and the term $\frac{1}{2}\bar{\psi}_{q}\psi_{q}\delta_{S}$, where $\delta_{ S}$ is the Dirac delta, ensures continuity on the surface $S$. For the case of the spherical bag of radius $R$, the argument of the functions $\Theta_{V}$ and $\delta_{ S}$ is $(R - r)$. The equations of motion are here obtained by means of the Euler-Lagrange equations \cite{Bjorken:1965zz} applied to the Lagrangian density in expression (\ref{lagrangian_density}).

Taking into account all the above considerations, the EoS for the MIT bag model takes the following form
\begin{equation}
\mathcal{E} = \frac{3}{\pi^{2}} \sum_{q} \int^{K_{F_{q}}}_{0}k^2 (m_{q}^2 +k^2)^{1/2}dk + B,
\label{mitenergdens}
\end{equation}
and
\begin{equation}
P = \frac{1}{\pi^{2}} \sum_{q} \int^{K_{F_{q}}}_{0}\frac{k^4}{(m_{q}^2 +k^2)^{1/2}}dk - B,
\label{mitpress}
\end{equation}
which are the expressions for energy density and pressure, respectively. Besides, $K_{F_{q}}$ is the Fermi moment. 

The baryonic number density ($\rho$) is given by the equation 
\begin{equation}
\rho = 
\sum_{q}\frac{1}{3} \rho_q =
\sum_{q}\frac{1}{3}\frac{(K_{F_{q}})^{3}}{\pi^{2}},
\label{barionic_number}
\end{equation}
where $\rho_q$ is the density of the quark of flavour $q$. Equations (\ref{mitenergdens})-(\ref{barionic_number}) are sufficient to describe quark matter. Whenever stellar matter is considered, chemical $\beta$ equilibrium and charge neutrality equations have to be imposed and hence, the inclusion of leptons (generally electrons and muons) is necessary. The relations between the chemical potentials of the different particles are given by the $\beta$-equilibrium conditions, 
\begin{equation}
\mu_s=\mu_d=\mu_u+\mu_e,
\qquad
\mu_e=\mu_\mu.
\label{qch}
\end{equation}

Moreover, for charge neutrality we must impose
\begin{equation}
\rho_e+\rho_\mu=\frac{1}{3}(2\rho_u-\rho_d-\rho_s).
\label{conditionsq}
\end{equation}
The electron and muon densities read
\begin{equation}
\rho_l={K_{F_l}^3}/{3\pi^2},
\end{equation}
where $K_{F_{l}}$ is the Fermi momentum for leptons. The energy density and pressure for the leptons are given, respectively, by
\begin{equation}
\mathcal{E}_l = \frac{3}{\pi^{2}}  \int^{K_{F_{l}}}_{0}k^2 (m_{l}^2 +k^2)^{1/2}dk,
\label{leptonener}
\end{equation}
\begin{equation}\label{leptonpress}
P = \frac{1}{3 \pi^{2}} \int^{K_{F_{l}}}_{0}\frac{k^4}{(m_{l}^2 +k^2)^{1/2}}dk,
\end{equation}
where the electron mass is taken as $0.511\, \rm MeV$ and the muon mass is $105.66\, \rm MeV$. We use a value for the bag constant equal to $B^{1/4} = 148\, \rm MeV$, which satisfies the stability window for this model according to the results shown in Ref.~\cite{Torres2013}.

\subsubsection{Hadronic matter}\label{sub_hadronic}

The relativistic model used here to describe the hadronic matter is a rather generalized version of the quantum hadrodynamics (QHD) \cite{livrowalecka, walecka, bogutabodmer}, which is based on a relativistic mean--field theory and describes the baryon interaction through the exchange of scalar and vector mesons, known as nonlinear Walecka model (NLWM). 

Since the introduction of the $\sigma-\omega$ model \cite{Walecka-74}, several relativistic hadronic models have been used with great success to describe infinite nuclear matter, finite nuclei and stellar matter properties \cite{Dutra2014}. Despite the good description of the binding energy, other important properties, such as incompressibility and effective mass of nucleons, are not obtained with satisfactory values. This problem was circumvented by Boguta and Bodmer in Ref.~\cite{bogutabodmer} introducing self-interaction terms, cubic and quartic, in the scalar field. Likewise, to deal with asymmetric systems in numbers of protons and neutrons, the vector-isovector meson $\rho$ was introduced, and to adjust other properties such as symmetry energy and the fact that protons and neutrons have slightly different masses, other mesons and interactions were being included, resulting in a huge number of generalizations and proposed parameterizations.

The complete Lagrangian density for the NLWM that describes baryons interacting among each other by exchanging scalar-isoscalar ($\sigma$), vector-isoscalar ($\omega$),
vector-isovector ($\rho$) and scalar-isovector ($\delta$) mesons is given by \cite{Agrawal2010,Dutra2014,Menezes2021}

\begin{widetext}
\begin{align}
\mathcal{L}_{\rm NLWM}  =&\ \overline{\psi}(i\gamma^\mu\partial_\mu - M)\psi 
+ g_\sigma\sigma\overline{\psi}\psi 
- g_\omega\overline{\psi}\gamma^\mu\omega_\mu\psi 
- \frac{g_\rho}{2}\overline{\psi}\gamma^\mu\vec{\rho}_\mu\cdot \vec{\tau}\psi
+ g_\delta\overline{\psi}\vec{\delta}\cdot \vec{\tau}\psi \nonumber \\ &+ \frac{1}{2}(\partial^\mu \sigma \partial_\mu \sigma 
- m^2_\sigma\sigma^2) - \frac{A}{3}\sigma^3 - \frac{B}{4}\sigma^4 - \frac{1}{4}F^{\mu\nu}F_{\mu\nu} 
+ \frac{1}{2}m^2_\omega\omega_\mu\omega^\mu 
+ \frac{C}{4}(g_\omega^2\omega_\mu\omega^\mu)^2 - \frac{1}{4}\vec{B}^{\mu\nu}\vec{B}_{\mu\nu} 
\nonumber \\ &+ \frac{1}{2}m^2_\rho\vec{\rho}_\mu \cdot \vec{\rho}^\mu + \frac{1}{2}(\partial^\mu\vec{\delta}\partial_\mu\vec{\delta}
- m^2_\delta\vec{\delta}^2) + g_\sigma g_\omega^2\sigma\omega_\mu\omega^\mu
\left(\alpha_1+\frac{1}{2}{\alpha_1}'g_\sigma\sigma\right)
+ g_\sigma g_\rho^2\sigma\vec{\rho}_\mu \cdot \vec{\rho}^\mu
\left(\alpha_2+\frac{1}{2}{\alpha_2}'g_\sigma\sigma\right) 
\nonumber \\
&+ \frac{1}{2}{\alpha_3}'g_\omega^2 g_\rho^2\omega_\mu\omega^\mu
\vec{\rho}_\mu\cdot \vec{\rho}^\mu.
\end{align}
\end{widetext}
In this Lagrangian density, $m_{i}$ represents the meson masses, with $i=\sigma,\omega,\rho,\delta$, and $g_{i}$ stands for the coupling constant of the interaction of the $i$ meson field with the baryonic field $\psi$. The antisymmetric field tensors $F_{\mu\nu}$ and $\vec{B}_{\mu\nu}$ are given by 
$F_{\mu\nu}=\partial_\nu\omega_\mu-\partial_\mu\omega_\nu$
and $\vec{B}_{\mu\nu}=\partial_\nu\vec{\rho}_\mu-\partial_\mu\vec{\rho}_\nu - g_\rho (\vec{\rho}_\mu \times \vec{\rho}_\nu)$. The coefficients $\gamma_{\mu}$ and $\vec{\tau}$ are the Dirac gamma matrices and Pauli matrices for the isospin, respectively. Finally, $M$ is the nucleon mass.

The equations of motion that describe the entire dynamics
of the system are obtained via Euler-Lagrange equations \cite{Bjorken:1965zz} and rotational and translational invariance are assumed. The result is coupled nonlinear equations whose solution is an extremely complicated task, even numerically. One approximation that can be made is the relativistic mean field (RMF) approximation, where the meson fields are treated as classical fields, and the substitution below is performed, namely
\begin{eqnarray}
\sigma\rightarrow \left<\sigma\right>\equiv\sigma_0, \quad \\
\omega_\mu\rightarrow \left<\omega_0\right>\equiv\omega_0, \quad \\
\vec{\rho}_\mu\rightarrow \left<\vec{\rho}_0\right>\equiv \bar{\rho}_{0(3)}, \quad \\
\vec{\delta}\rightarrow\,\,<\vec{\delta}>\equiv\delta_{(3)},
\label{meanfield}
\end{eqnarray}
and the equations of motion are given in the following expressions:
\begin{widetext}
\begin{align}
m^2_\sigma\sigma_0 &= g_\sigma\rho_s - A\sigma_0^2 - B\sigma_0^3 + g_\sigma g_\omega^2\omega_0^2(\alpha_1+{\alpha_1}'g_\sigma\sigma)
+g_\sigma g_\rho^2\bar{\rho}_{0(3)}^2(\alpha_2+{\alpha_2}'g_\sigma\sigma) ,  \label{sigmaacm}  \\
m_\omega^2\omega_0 &= g_\omega\rho - Cg_\omega(g_\omega \omega_0)^3 - g_\sigma g_\omega^2\sigma_0\omega_0(2\alpha_1+{\alpha_1}'g_\sigma\sigma_0) - {\alpha_3}'g_\omega^2 g_\rho^2\bar{\rho}_{0(3)}^2\omega_0 , \label{omegaacm}  \\
m_\rho^2\bar{\rho}_{0(3)} &= \frac{g_\rho}{2}\rho_3 
-g_\sigma g_\rho^2\sigma_0\bar{\rho}_{0(3)}(2\alpha_2+{\alpha_2}'g_\sigma\sigma_0) -{\alpha_3}'g_\omega^2 g_\rho^2\bar{\rho}_{0(3)}\omega_0^2 ,  \label{rhoacm} \\
m_\delta^2\delta_{(3)} &= g_\delta\rho_{s3},  \label{deltaacm}
\end{align}
\end{widetext}
and
\begin{equation}\label{diracacm}
 [ i\gamma^\mu \partial_\mu -\gamma^0 V_\tau  - (M+S_\tau)] \psi = 0. 
\end{equation}

The scalar and vector densities are given by
\begin{equation}
  \rho_s =\left<\overline{\psi}\psi\right>={\rho_s}_p+{\rho_s}_n,
\end{equation}
\begin{equation}\label{rhos_tot}
\rho_{s3}=\left<\overline{\psi}{\tau}_3\psi\right>={\rho_s}_p-{\rho_s}_n,
\end{equation}
\begin{equation}
  \rho =\left<\overline{\psi}\gamma^0\psi\right> = \rho_p + \rho_n,
\end{equation}
\begin{equation}\label{rho_tot}
\rho_3=\left<\overline{\psi}\gamma^0{\tau}_3\psi\right> = \rho_p - \rho_n=(2y_p-1)\rho,  
\end{equation}
with
\begin{equation}\label{rhospn}
{\rho_s}_{p,n} = \frac{\gamma M^*_{p,n}}{2\pi^2}\int_0^{{k_F}_{p,n}}
\frac{k^2dk}{\sqrt{k^2+M^{*2}_{p,n}}} ,
\end{equation}\
\begin{equation}\label{rhopn}
\rho_{p,n} = \frac{\gamma}{2\pi^2}\int_0^{{k_F}_{p,n}}k^2dk =
\frac{\gamma}{6\pi^2}{k_F^3}_{p,n},
\end{equation}
\begin{equation}
V_{\tau} =g_\omega\omega_0 +
\frac{g_\rho}{2}\bar{\rho}_{0(3)}\tau_3 ,  \quad
S_{\tau} =-g_\sigma\sigma_0 -g_\delta\delta_{(3)}\tau_3,
\end{equation}
with $\tau_3 = 1$ and $\tau_3 = -1$ for protons and neutrons, respectively. Here, $\gamma$ is the spin degeneracy. Besides, the proton and neutron effective masses are given by
\begin{equation}
    M^{*}_{p} = M - g_{\sigma}\sigma_{0}-g_{\delta}\delta_{(3)}, \quad M^{*}_{n} = M - g_{\sigma}\sigma_{0}+g_{\delta}\delta_{(3)}.
\end{equation}

After some analytical procedures, the expressions for energy density and pressure are obtained \cite{Dutra2014,Menezes2021}

\begin{widetext}
\begin{align}
\mathcal{E} =&\ \frac{1}{2}m^2_\sigma\sigma_0^2 
+ \frac{A}{3}\sigma_0^3 + \frac{B}{4}\sigma_0^4 - \frac{1}{2}m^2_\omega\omega_0^2 
- \frac{C}{4}(g_\omega^2\omega_0^2)^2 - \frac{1}{2}m^2_\rho\bar{\rho}_{0(3)}^2
+g_\omega\omega_0\rho+\frac{g_\rho}{2}\bar{\rho}_{0(3)}\rho_3
\nonumber \\
&+ \frac{1}{2}m^2_\delta\delta^2_{(3)} - g_\sigma g_\omega^2\sigma\omega_0^2
\left(\alpha_1+\frac{1}{2}{\alpha_1}'g_\sigma\sigma_0\right) 
- g_\sigma g_\rho^2\sigma\bar{\rho}_{0(3)}^2 
\left(\alpha_2+\frac{1}{2}{\alpha_2}' g_\sigma\sigma_0\right) \nonumber \\
&- \frac{1}{2}{\alpha_3}'g_\omega^2 g_\rho^2\omega_0^2\bar{\rho}_{0(3)}^2
+ \mathcal{E}_{\mbox{\tiny kin}}^p + \mathcal{E}_{\mbox{\tiny kin}}^n,
\label{denerg}
\end{align}
\end{widetext}
with
\begin{eqnarray}
\mathcal{E}_{\mbox{\tiny kin}}^{p,n}&=&\frac{\gamma}{2\pi^2}\int_0^{{k_F}_{p,n}}k^2
(k^2+M^{*2}_{p,n})^{1/2}dk,
\label{decinnlw}
\end{eqnarray}\\
and pressure:
\begin{align}
P =& - \frac{1}{2}m^2_\sigma\sigma_0^2 - \frac{A}{3}\sigma_0^3 -\frac{B}{4}\sigma_0^4 + \frac{1}{2}m^2_\omega\omega_0^2 
+ \frac{C}{4}(g_\omega^2\omega_0^2)^2  \nonumber \\
&+ \frac{1}{2}m^2_\rho\bar{\rho}_{0(3)}^2
+ \frac{1}{2}{\alpha_3}'g_\omega^2 g_\rho^2\omega_0^2\bar{\rho}_{0(3)}^2 -\frac{1}{2}m^2_\delta\delta^2_{(3)}  \nonumber \\
&+ g_\sigma g_\omega^2\sigma_0\omega_0^2
\left(\alpha_1+\frac{1}{2}{\alpha_1}'g_\sigma\sigma_0\right)  \nonumber \\
&+ g_\sigma g_\rho^2\sigma\bar{\rho}_{0(3)}^2 
\left(\alpha_2+\frac{1}{2}{\alpha_2}' g_\sigma\sigma\right) + P_{\mbox{\tiny kin}}^p + P_{\mbox{\tiny kin}}^n , \
\label{pressure}
\end{align}
where
\begin{equation}
P_{\mbox{\tiny kin}}^{p,n} = 
\frac{\gamma}{6\pi^2}\int_0^{{k_F}_{p,n}}\frac{k^4dk}{(k^2+M^{*2}_{p,n})^{1/2}}.
\end{equation}

At this stage, charge neutrality and chemical equilibrium conditions need to be implemented, which depend on the inclusion of leptons. These conditions read
\begin{align}\label{conditionH}
    \mu_p &= \mu_n -\mu_e,  &  \mu_e &=\mu_\mu,  &  \rho_p &= \rho_e + \rho_\mu ,
\end{align}
where, the leptonic expressions for energy density, pressure and density are the same given in \ref{sub_quarks}.

Of the many possible parameterizations of the QHD model, we chose the IU-FSU parameterization proposed by Piekarewicz and authors in \cite{iufsu}. In addition to satisfying the constraints investigated in \cite{Dutra2014,Dutra2015}, the IU-FSU model also satisfactorily reproduces the recent astrophysical constraint from the observation of GW170817 \cite{Lourenco2018}. 

Finally, to describe the star outer crust, we use the full BPS EoS \cite{bps}.


\section{Dynamical stability}\label{Sect3}

The solution of the modified TOV equations (\ref{TOV1})-(\ref{TOV3}) provides a family of equilibrium configurations, but such an equilibrium state may be stable or unstable with respect to a radial perturbation. As mentioned in the introduction, a technique widely used in the literature to identify the onset of instability is the $M(\rho_c)$ method, that is, the first maximum in the mass-central density curve marks the boundary between stable and unstable stars. However, this method is very simple and provides a necessary but not sufficient condition for stability analysis along the sequence of equilibrium configurations \cite{Glendenning2000, Horvat2011}. A more rigorous analysis of the radial stability of a compact star involves calculating the eigenfrequencies of the normal radial modes of pulsation (which may grow or be damped exponentially). In that regard, here we are going to derive for the first time the differential equations that govern the adiabatic radial oscillations within the context of Rastall gravity. With this in mind, we will deal with radial motions so that the stellar system maintains its spherical symmetry. Furthermore, in the adiabatic approximation, the heat transfer can be ignored during the dynamics of small perturbations.

We regard small deviations from the hydrostatic equilibrium where a fluid element located at $r$ in the unperturbed configuration (characterized by zero velocities) is displaced to $r+ \xi(t,r)$ in the perturbed system, and hence we can define $v = \partial\xi/\partial t$. Moreover, the variable $h$ (representing any metric or fluid quantity) can be written as $h(t,r) = h_0(r)+ \delta h(t,r)$, where $h_0$ is the background solution, $\delta h$ is the Eulerian perturbation and $\vert \delta h/h_0 \vert \ll 1$. Henceforth we will only maintain first-order terms in all perturbations. This means that we can write the four-velocity (\ref{4Velocity}) as $u^\mu = (e^{-\psi}, e^{-\psi_0}v, 0, 0)$, and the energy-momentum tensor (\ref{EMtensor}) takes the form \cite{Pretel2021JCAP} 
\begin{equation}
T_\mu^{\ \nu}=  \begin{pmatrix}
  -\rho & -(\rho_0 + p_0)v & 0 & 0 \\
  (\rho_0+p_0)ve^{2\lambda_0- 2\psi_0} & p & 0 & 0 \\
  0 & 0 & p & 0 \\
  0 & 0 & 0 & p
\end{pmatrix} .
\end{equation}

Although the procedure to derive the radial oscillation equations is a bit cumbersome, here we are going to summarize the main steps to obtain them. The approach is analogous to that carried out by Chandrasekhar in conventional GR \cite{ChandrasekharApj}. Taking into account the static background equations (\ref{ModEq1})-(\ref{ModEq4}), the perturbation of the field equations (\ref{FEq1}), (\ref{FEq2}) and (\ref{FEq4}) together
with the non-conservative Eq.~(\ref{4diverA}) leads to
\begin{equation}\label{PerturEq1}
        \delta\lambda = -\xi(\psi'_0 + \lambda'_0) ,
    \end{equation}
\begin{widetext}
    \begin{equation}\label{PerturEq2}
        (\rho_0+ p_0)\frac{\partial}{\partial r}(\delta\psi) = \left[ (1- 3\beta)\delta p + \beta\delta\rho - (\rho_0 + p_0)\left( \frac{1}{r} + 2\psi'_0 \right)\xi \right](\psi'_0 + \lambda'_0) , 
    \end{equation}
    \begin{equation}\label{PerturEq3}
        \delta\rho = -\frac{3\beta}{1-\beta}(\delta p + \xi p'_0) - \xi\rho'_0 - \left[ \frac{\rho_0 + p_0}{1- \beta} \right] \frac{e^{\psi_0}}{r^2}\frac{\partial}{\partial r}\left(r^2\xi e^{-\psi_0}\right) ,
    \end{equation}
    \begin{equation}\label{PerturEq4}
        (\rho_0+ p_0)e^{2\lambda_0 - 2\psi_0}\frac{\partial v}{\partial t} + \frac{\partial}{\partial r}(\delta p) + (\rho_0 + p_0)\frac{\partial}{\partial r}(\delta\psi) + \psi'_0(\delta\rho + \delta p) = \beta\frac{\partial}{\partial r}(3\delta p - \delta\rho) .
    \end{equation}
\end{widetext}
Now let us assume that all perturbations have a harmonic time dependence, namely, $\xi(t,r) = \chi(r)e^{i\omega t}$ and $\delta h(t,r) = \delta h(r)e^{i\omega t}$, where $\omega$ is the eigenfrequency of the radial vibrations to be determined. This assumption allows us to cancel the exponential factors in each linearized equation and hence we can obtain time-independent equations. This means that all equations are now in terms of the amplitudes $\delta h(r)$ and quantities of the static background, and thus we can also take out the subscript zero. In view of Eq.~(\ref{PerturEq2}), the linearized non-conservation equation (\ref{PerturEq4}) becomes 
\begin{align}\label{NonConserEq1}
    &\omega^2(\rho+ p)e^{2\lambda - 2\psi}\chi = (1- 3\beta)\frac{d(\delta p)}{dr} + \beta\frac{d(\delta\rho)}{dr}  \nonumber  \\
    & + \left[ (2- 3\beta)\psi' + (1- 3\beta)\lambda' \right]\delta p + \left[ (1+ \beta)\psi' + \beta\lambda' \right]\delta\rho  \nonumber  \\
    &- (\rho+ p)\left( \frac{1}{r}+ 2\psi' \right)(\psi' + \lambda')\chi .
\end{align}

If the pressure is a function only of the energy density (this is, $p = p(\rho)$), we can write $\delta p = (dp/d\rho) \delta\rho$. As a consequence, by means of Eq.~(\ref{PerturEq3}), we obtain 
\begin{equation}\label{DeltapEq1}
    \delta p = -\chi p' - \frac{1}{1-\beta}\left[ 3\beta\Delta p\frac{dp}{d\rho} + \Gamma p\frac{e^\psi}{r^2}\frac{d}{dr}(r^2\chi e^{-\psi}) \right] ,
\end{equation}
where $\Delta p = \delta p + \chi p'$ is the Lagrangian perturbation of the pressure, and $\Gamma$ is the adiabatic index at constant entropy:
\begin{equation}
    \Gamma = \left( 1 + \frac{\rho}{p} \right)\frac{dp}{d\rho} .
\end{equation}

Note that Eq.~(\ref{DeltapEq1}) can be rewritten in a more compact form for the Lagrangian perturbation $\Delta p$, namely
\begin{equation}\label{DeltapEq2}
    \Delta p = -\frac{\Gamma p}{\mathcal{J}} \frac{e^\psi}{r^2}\frac{d}{dr}(r^2\chi e^{-\psi}) ,
\end{equation}
where $\mathcal{J}$ has been defined as 
\begin{equation}
    \mathcal{J} \equiv 1 - \beta + 3\beta\frac{dp}{d\rho} .
\end{equation}

Furthermore, Eq.~(\ref{PerturEq3}) can be rewritten as 
\begin{equation}\label{deltaRhoEq}
    \delta\rho = -\frac{3\beta}{1- \beta}\delta p - \frac{\mathcal{Q}}{1- \beta} , 
\end{equation}
with $\mathcal{Q}$ being given by
\begin{align}\label{QEq}
    \mathcal{Q} &\equiv \frac{d}{dr}\left[ (\rho+ p)\chi \right] + \frac{2}{r}(\rho+ p)\chi  \nonumber  \\
    &= \chi\rho' - \beta\chi(\rho' - 3p') - \left( \frac{\rho+ p}{\Gamma p} \right)\mathcal{J}\Delta p .
\end{align}

After substituting Eqs.~(\ref{DeltapEq1}) and (\ref{deltaRhoEq}) into (\ref{NonConserEq1}), and by using Eqs.~(\ref{FEq3}) and (\ref{ModEq3}), we have 
\begin{widetext}
  \begin{align}
    \omega^2(\rho + p)e^{2\lambda- 2\psi}\chi =&\  \frac{1-4\beta}{1- \beta} \bigg\lbrace (\Delta p)' + (2\psi'+ \lambda')\Delta p + (\rho +p)\left[ 8\pi p+ 8\pi\beta(\rho - 3p) \right]e^{2\lambda}\chi  \nonumber \\
    &- \chi(\rho+ p)\psi'^2 + \frac{4}{r}\chi p' + \beta\frac{d}{dr}[\chi(\rho' - 3p')] + 2\beta\chi(\rho'- 3p')\left( \frac{2}{r} + \psi' + \frac{\lambda'}{2} \right) \bigg\rbrace  \nonumber  \\
    &- \frac{\beta}{1- \beta}\left[ 3(\rho+ p)\left( \frac{1}{r} + 2\psi' \right)(\psi' + \lambda')\chi + \mathcal{Q}' + (5\psi'+ \lambda')\mathcal{Q} \right] ,
  \end{align}
and through Eq.~(\ref{QEq}), the last expression becomes 
  \begin{align}\label{DeltapPrimeEq}
    \mathcal{G}(\Delta p)' =&\ \chi\bigg\lbrace \frac{1-\beta}{1- 4\beta}(\rho +p)\left[ \omega^2e^{-2\psi} - 8\pi p - 8\pi\beta(\rho- 3p) \right]e^{2\lambda} - \frac{1-7\beta}{1-4\beta}\frac{4}{r}p' + \frac{1+5\beta}{1-4\beta}(\rho+ p)\psi'^2  \nonumber  \\
    &- \frac{\beta}{1-4\beta}\left[ \frac{4}{r}\rho' - \beta(\rho'- 3p')\left( \frac{16}{r} + 3\psi'+ 3\lambda' \right) - 3p'(\psi'+ \lambda') - 3(\rho+ p)\left( \lambda'+ \frac{2}{r} \right)\psi' \right] \bigg\rbrace  \nonumber  \\
    &- \Delta p \left\lbrace 2\psi'+ \lambda' + \frac{\beta}{1- 4\beta}\left[ (5\psi' + \lambda')\frac{\rho+ p}{\Gamma p}\mathcal{J} + \frac{d}{dr}\left( \frac{\rho+ p}{\Gamma p}\mathcal{J} \right) \right] \right\rbrace + \frac{3\beta}{1- 4\beta}\left[ \beta(\rho' - 3p') + p'\right]\chi' ,
  \end{align}
where we have defined 
\begin{equation}
    \mathcal{G} \equiv 1+ \frac{\beta}{1- 4\beta}\left( \frac{\rho+ p}{\Gamma p}\mathcal{J} \right) .
\end{equation}

Finally, it is convenient to introduce a new variable given by $\zeta \equiv \chi/r$. Thus, from Eqs.~(\ref{DeltapEq2}) and (\ref{DeltapPrimeEq}) we obtain two first-order differential equations governing the adiabatic radial pulsations for a compact star in Rastall gravity:
\begin{align}
    \frac{d\zeta}{dr} =& -\frac{1}{r}\left[ 3\zeta + \frac{\mathcal{J}}{\Gamma p}\Delta p \right] + \psi'\zeta ,  \label{RadialOsEq1}  \\
    \mathcal{G}\frac{d(\Delta p)}{dr} =&\ \zeta\bigg\lbrace \frac{1-\beta}{1- 4\beta}(\rho +p)\left[ \omega^2e^{-2\psi} - 8\pi p - 8\pi\beta(\rho- 3p) \right]re^{2\lambda} - 4\left(\frac{1-7\beta}{1-4\beta}\right)p' + \frac{1+5\beta}{1-4\beta}(\rho+ p)r\psi'^2  \nonumber  \\
    &+ \frac{\beta r}{1-4\beta}\left[ \beta(\rho'- 3p')\left( \frac{19}{r}+ 3\psi' + 3\lambda' \right)+ 3p'\left( \frac{1}{r}+ \psi' + \lambda' \right) - \frac{4}{r}\rho' + 3(\rho+ p)\left( \lambda'+ \frac{2}{r} \right)\psi' \right] \bigg\rbrace  \nonumber  \\
    &- \Delta p \left\lbrace 2\psi'+ \lambda' + \frac{\beta}{1- 4\beta}\left[ (5\psi' + \lambda')\frac{\rho+ p}{\Gamma p}\mathcal{J} + \frac{d}{dr}\left( \frac{\rho+ p}{\Gamma p}\mathcal{J} \right) \right] \right\rbrace + \frac{3\beta r}{1- 4\beta}\left[ \beta(\rho' - 3p') + p'\right]\zeta' , \label{RadialOsEq2} 
\end{align}

\end{widetext}
and such equations reduce to the corresponding radial oscillation equations in standard GR when $\beta= 0$, see for instance Refs.~\cite{Gondek1997, Flores2010, Pereira2018, Pretel2020MNRAS}. Note that Eq.~(\ref{RadialOsEq1}) has a trivial coordinate singularity at the center and hence the coefficient of $1/r$ term must vanish at $r= 0$. Furthermore, the condition $p(r_{\rm sur})= 0$ leads to imposing another boundary condition for the Lagrangian perturbation of pressure at the surface. 
Consequently, the system of equations (\ref{RadialOsEq1}) and (\ref{RadialOsEq2}) will be solved by means of the following boundary conditions
\begin{align}
    \Delta p &= -3\frac{\Gamma p\zeta}{\mathcal{J}}  \qquad \  \text{as}  \qquad  \  r\rightarrow 0 ,  \label{BCRadO1}  \\
    \Delta p &= 0  \qquad \  \text{as}  \qquad  \  r\rightarrow r_{\rm sur} .  \label{BCRadO2}
\end{align}


\section{Numerical results}\label{Sect4}

In this section, we present and discuss our numerical results. The essential ingredients of nuclear physics for astrophysical calculations are the appropriate EoSs. Such equations of state will be the input to the modified hydrostatic equilibrium and radial oscillation equations in Rastall gravity. For quark matter (\ref{sub_quarks}) we used the MIT bag model \cite{Chodos1974}, in which the energy density and pressure are described by the expressions (\ref{mitenergdens}) and (\ref{mitpress}). Besides that, for hadronic matter (\ref{sub_hadronic}) describing neutron stars, we used the IU-FSU parameterization \cite{iufsu} of the Walecka model with nonlinear terms \cite{livrowalecka, walecka, bogutabodmer}, where energy density and pressure are given by Eqs.~(\ref{denerg}) and (\ref{pressure}), respectively.

\subsection{Hydrostatic equilibrium}

\begin{figure*}
 \includegraphics[width=8.6cm]{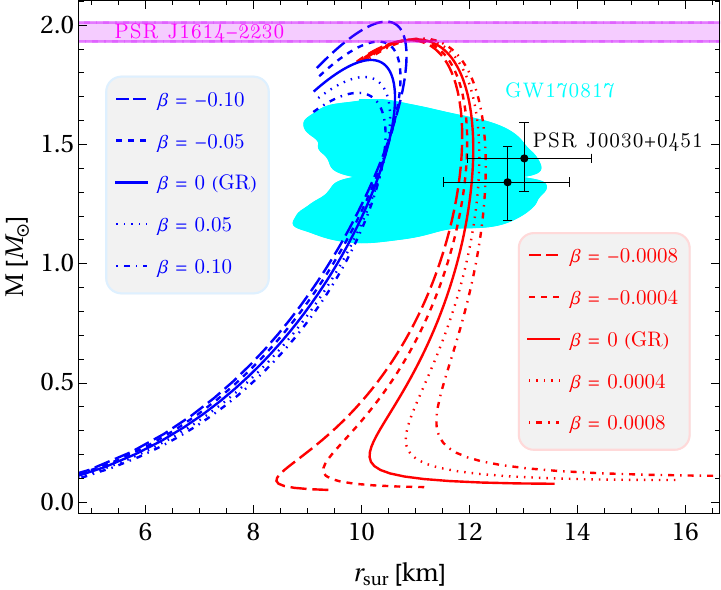}
 \includegraphics[width=8.625cm]{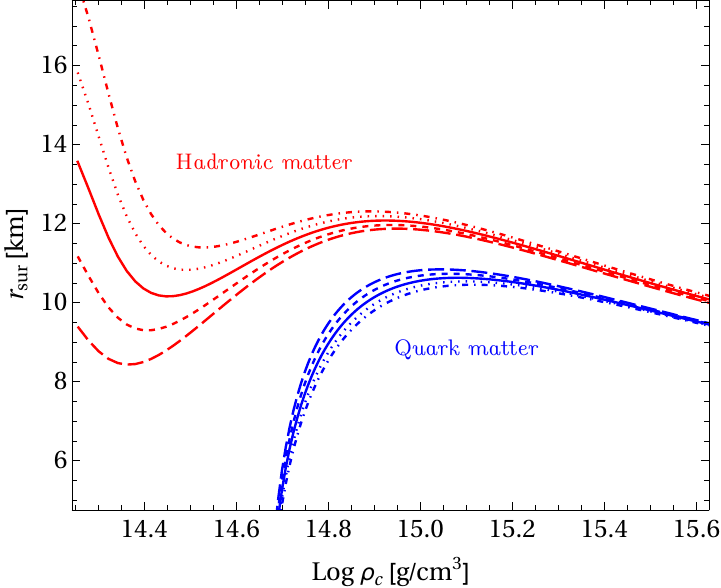}
 \caption{\label{figure1} Left panel: Mass-radius diagram for quarks stars (blue lines) and neutron stars (red lines) as predicted by Rastall gravity for different values of $\beta$. The magenta horizontal band represents the observational measurement for the millisecond pulsar J1614-2230 reported in Ref.~\cite{Demorest2010}. The NICER measurements for PSR J0030+0451 are shown by black dots with their respective error bars \cite{Miller:2019cac, Riley:2019yda}, and the cyan area stands for the mass-radius constraint from the GW170817 event. Right panel: Radius versus central density relation. In the case of neutron stars, the more substantial deviations for $r_{\rm sur}$ from GR ($\beta= 0$) take place in the low-mass region, whereas for large masses (close to the maximum mass) the changes are very slight due to the Rastall parameter. Meanwhile, for quark matter, the radius undergoes little pronounced modifications and the maximum-mass values reveal relevant changes when $\vert\beta\vert$ takes larger values than in the case of hadronic matter. }  
\end{figure*}

Our first task is to obtain the metric variables and the thermodynamic quantities as functions of the radial coordinate. To do so, the stellar structure equations (\ref{TOV1})-(\ref{TOV3}) with boundary conditions (\ref{BConditions}) will be numerically integrated from the origin up to the surface for a given equation of state $p(\rho)$. As usual, the radius of the star is determined when the pressure vanishes (i.e., $p(r_{\rm sur})= 0$), and the total gravitational mass is given by $M \equiv m(r_{\rm sur})$. In the left panel of Fig.~\ref{figure1}, we plot the total mass versus radius for quark (blue lines) and hadronic (red lines) matter. Here we have considered values of $\beta$ for which appreciable changes can be observed in the mass-radius diagram. The radius of neutron stars undergoes considerable deviations from GR in the low-mass region. This can be better observed when we plot the radius as a function of the central density, see the right panel of the same figure. Positive (negative) values of $\beta$ result in larger (smaller) radii with respect to their pure general relativistic counterpart. Nonetheless, it should be noted that the changes in the maximum-mass values are insignificant.

On the other hand, noticeable changes in the most basic properties of a quark star are observed if we use larger values of $\vert\beta\vert$ than in the case of hadronic matter. The main consequence of the Rastall term on quark stars is a significant increase (decrease) in the maximum mass for negative (positive) values of $\beta$. Furthermore, according to the right plot of Fig.~\ref{figure1}, we observe that a negative value of $\beta$ increases the radius of a quark star, for a fixed central density. Remarkably, this behavior is contrary to the case of neutron stars.

Our numerical results provide a realistic description of compact stars in Rastall gravity in the sense that they satisfy the observational measurements. Specifically, for both equations of state, the set of values adopted for $\beta$ satisfies the mass-radius constraint from the GW170817 event (see the filled cyan region). Moreover, the range $\beta \in [-0.10, -0.05]$ consistently describes the millisecond pulsar J1614-2230 for the quark matter EoS. For the pulsar PSR J0030+0451, which was observed using the Neutron Star Interior Composition Explorer (NICER), we notice that positive values of $\beta$ are more suitable for its description using a hadronic matter EoS.

\begin{figure}
 \includegraphics[width=8.55cm]{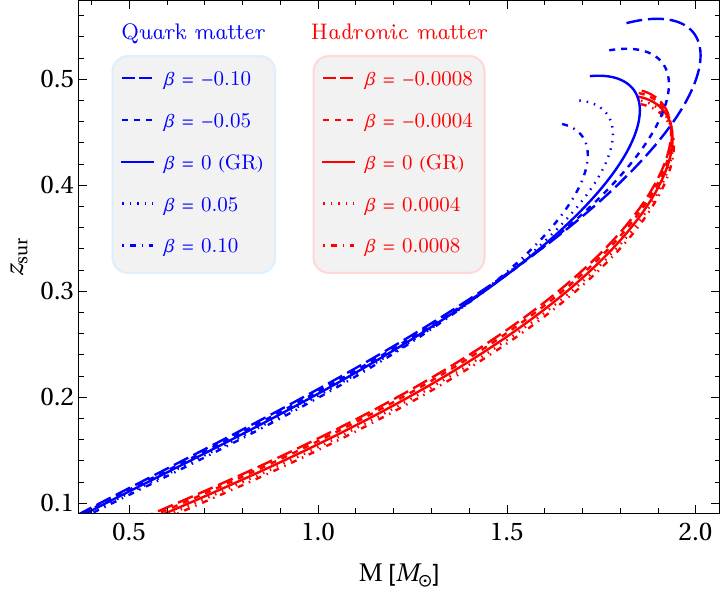}
 \caption{\label{figure2} Gravitational redshift of light emitted at the stellar surface as a function of the total gravitational mass for the same values of $\beta$ as considered in Fig.~\ref{figure1}. The redshift exhibits significant changes only in the high-mass region for quark stars. }  
\end{figure}

Another interesting quantity that in principle can be observed is the gravitational redshift given by Eq.~(\ref{RedShiftEq}). Figure \ref{figure2} shows the redshift of light emitted at the surface of each star as a function of gravitational mass for the same set of values of $\beta$ as in Fig.~\ref{figure1}. It can be seen that the maximum value of $z_{\rm sur}$ increases substantially with decreasing $\beta$, while the changes are smaller for hadronic matter.

\begin{figure}
 \includegraphics[width=8.55cm]{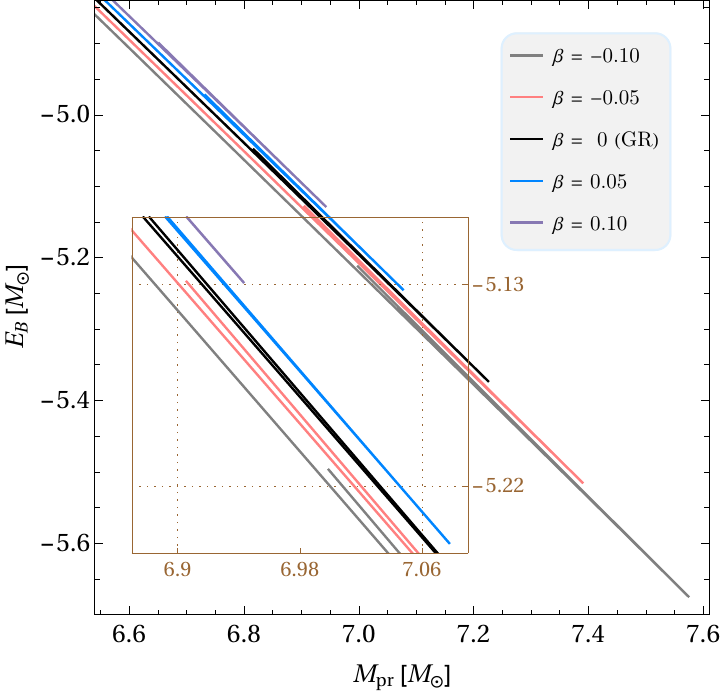}
 \caption{\label{BindEner1} Binding energy $E_B$ as a function of the proper mass $M_{\rm pr}$ of quark stars in Rastall gravity. Observe the formation of a cusp when the binding energy becomes minimal (see also Fig.~\ref{BindEner2} to visualize these minima as a function of the central energy density). }  
\end{figure}

Following Refs.~\cite{Doneva2013, PRETEL2024PDU, Doneva2021, Jimenez2022}, the gravitational binding energy $E_B$ can be plotted as a function of the proper mass $M_{\rm pr}$ in order to analyze the stellar stability of the equilibrium configurations. The proper mass (also known as baryon mass) of a compact star is defined as
\begin{equation}\label{EqMpr}
    M_{\rm pr} = 4\pi m_B\int_0^{r_{\rm sur}} e^{\lambda(r)}r^2n_B(r)dr ,
\end{equation}
with $n_B(r)$ being the baryon number density function and $m_B$ is the neutron mass which is chosen to agree with different stellar evolution models requiring the conservation of baryon number \cite{Jimenez2022}. The difference between the total mass distribution of standard matter $M_\rho \equiv m_\rho(r_{\rm sur})$ and the proper mass gives us the binding energy, namely $E_B = M_\rho - M_{\rm pr}$ \cite{Pretel2021JCAP}. For quark matter EoS, the binding energy versus proper mass relation is shown in Fig.~\ref{BindEner1}, where we can appreciate the emergence of a cusp for the different values of $\beta$.

\begin{figure}
 \includegraphics[width=8.5cm]{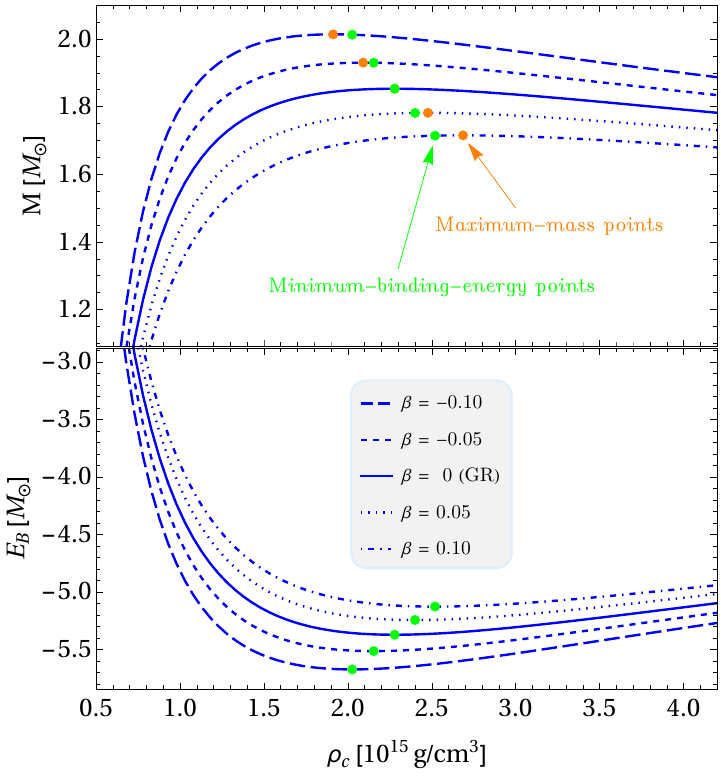}
 \caption{\label{BindEner2} Upper panel: Mass-central density relation. Lower panel: Variation of the gravitational binding energy with respect to $\rho_c$ for different values of $\beta$ using the quark matter EoS. The full orange and green circles indicate the maximum-mass points and the minimum-binding-energy points, respectively. See also the left plot of Fig.~\ref{SFFigMIT}, where the same values of central density corresponding to the minimum of $E_B$ have been located in the inset. }  
\end{figure}

As already mentioned in the introduction, a turning point from stability to instability occurs when $dM/d\rho_c =0$ according to the $M(\rho_c)$ method. In other words, the stable branch of the quark stars shown in the left plot of Fig.~\ref{figure1} must be located below the critical central density corresponding to the maximum-mass configuration (see the full orange circles in the upper panel of Fig.~\ref{BindEner2}). However, due to its simplicity, this criterion does not provide consistent results with the binding energy calculation. In particular, we find that the maximum-mass points and the minimum-binding-energy points do not coincide. This is because the Rastall term generates an extra mass contribution in Eq.~(\ref{Massfun2}). A more rigorous approach to investigate the radial stability of compact stars is by calculating the frequencies of the radial vibration modes which will be discussed below.

\begin{figure*}
 \includegraphics[width=8.56cm]{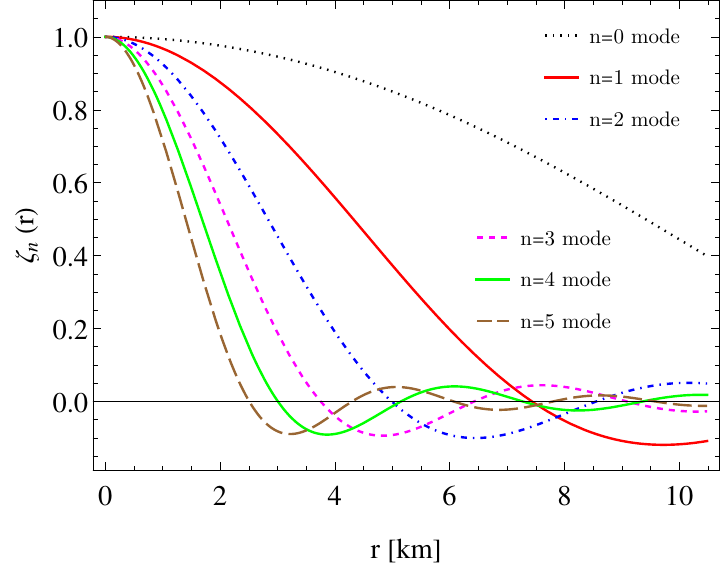}
 \includegraphics[width=8.78cm]{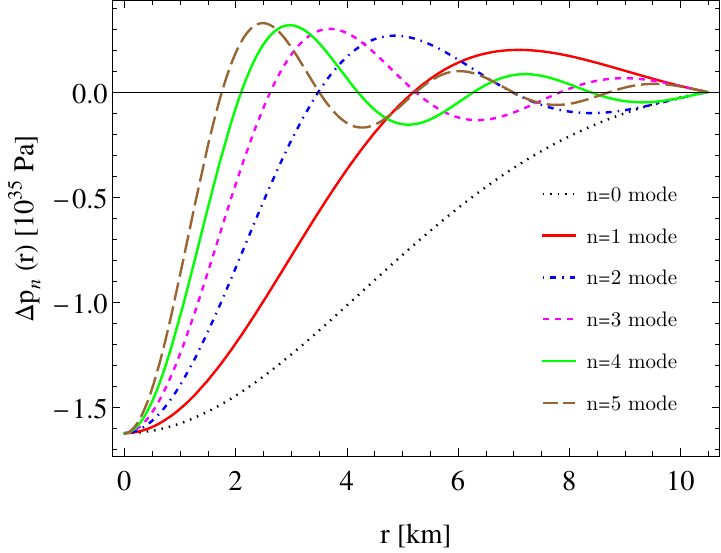}
 \caption{\label{EigenfuncFig} Eigenfunctions $\zeta_n(r)$ in the left panel and $\Delta p_n(r)$ in the right panel for the first six normal vibration modes as a function of the radial coordinate. The two plots correspond to a quark star in Rastall gravity with central density $\rho_c= 1.5\times 10^{15}\, \rm g/cm^3$ and $\beta = 0.05$. In can be observed that the solution corresponding to the $n$th oscillation mode has $n$ nodes inside the star, where $n=0$ describes the fundamental mode. Note also that the perturbations $\zeta_n(r)$ have been normalized at the origin, i.e. $\zeta (0) =1$, and the Lagrangian perturbation of the pressure $\Delta p_n(r)$ satisfies the boundary condition (\ref{BCRadO2}) at the surface. }  
\end{figure*}

\subsection{Radial pulsations}

Once the static background quantities (as well as the introduced variables $\mathcal{J}$ and $\mathcal{G}$) are determined by integrating the modified TOV equations, our second task is to solve the radial pulsation equations in order to investigate the dynamical stability of the stellar configurations presented in Fig.~\ref{figure1}. The system of differential equations (\ref{RadialOsEq1}) and (\ref{RadialOsEq2}), subject to the boundary conditions (\ref{BCRadO1}) and (\ref{BCRadO2}), constitutes a Sturm-Liouville eigenvalue problem for the squared frequencies. The corresponding solutions $\zeta(r)$ and $\Delta p(r)$ are the eigenfunctions associated to each eigenvalue $\omega^2$. In other words, there is an infinite number of solutions, but only specific values of $\omega^2$ which properly satisfy the boundary conditions are allowed.

As in the pure general relativistic case \cite{Gondek1997, Flores2010, Pereira2018, Pretel2020MNRAS}, here we use the shooting method to solve first-order differential equations (\ref{RadialOsEq1}) and (\ref{RadialOsEq2}), namely, we perform the integration for a set of test values $\omega^2$ obeying the condition (\ref{BCRadO1}). Moreover, we assume that the normalized eigenfunctions correspond to $\zeta(0)= 1$ at the stellar center, and we integrate up to the surface. The guessed values of $\omega^2$ that satisfy the boundary condition (\ref{BCRadO2}) will be the correct frequencies of the different radial vibration modes. In particular, for a specific value of the parameter $\beta$ and considering a quark star with central density $\rho_c = 1.5\times 10^{15}\, \rm g/cm^3$, the first six normal pulsations modes are shown in Fig.~\ref{EigenfuncFig}. Such configuration has a frequency spectrum $\omega_0^2< \omega_1^2< \cdots< \omega_n^2< \cdots$, where $n$ represents the number of nodes between the center and the stellar surface. The first eigenvalue corresponding to $n=0$ is the fundamental or nodeless mode (see the black dotted line in Fig.~\ref{EigenfuncFig}) because it has the lowest frequency, while the higher frequencies are called overtones or excited modes. So, the first excited mode $n=1$ (red curve) has a node, the second excited mode $n=2$ (blue curve) has two, and so forth. If any of these eigenvalues is negative (i.e., $\omega^2< 0$) for a particular stellar configuration, the frequency is purely imaginary and hence any radial perturbation of the star will generate a dynamical instability. On the other hand, the case $\omega^2> 0$ describes an oscillatory behavior and thereby corresponds to a stable star.

\begin{figure*}
 \includegraphics[width=8.56cm]{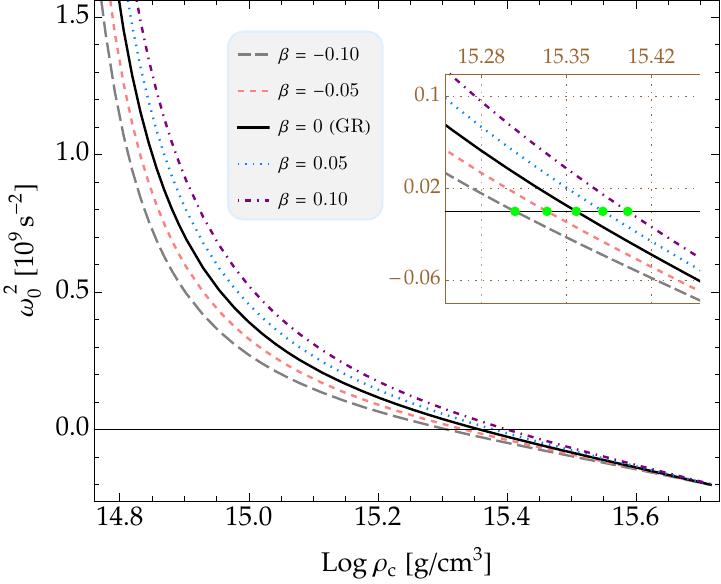}
 \includegraphics[width=8.63cm]{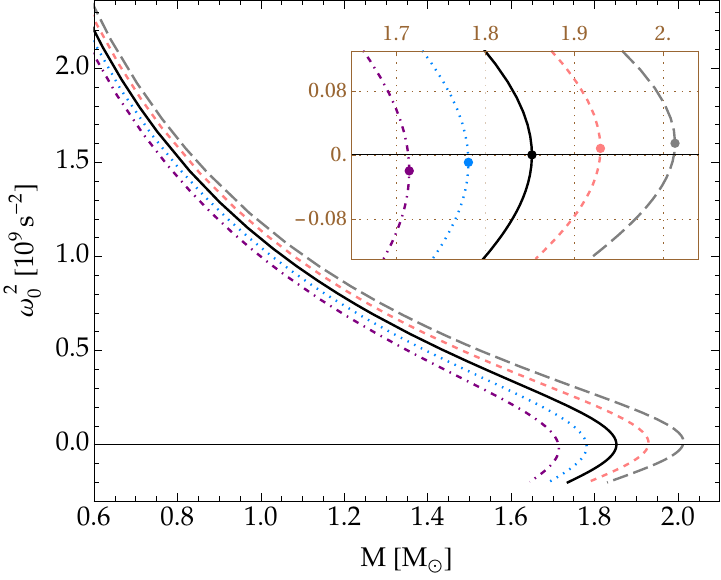}
 \caption{\label{SFFigMIT} Left panel: Squared frequency of the fundamental vibration mode as a function of the central density predicted by Rastall gravity for quark matter, where the different lines indicate different values of the parameter $\beta$. The inset magnifies the behavior of $\omega_0^2$ in the surroundings of its vanishing value, where the green points correspond precisely to the minimum-binding-energy points in Fig.~\ref{BindEner2}. Right panel: Squared frequency of the fundamental oscillation mode versus gravitational mass. Note that, when $\beta\neq 0$, the maximum-mass values do not correspond to $\omega_0^2= 0$. This means that the conventional condition $dM/d\rho_c =0$ to indicate the onset of instability is no longer valid in Rastall gravity. Stars cease to be stable before the maximum-mass configuration for positive values of $\beta$, while stable stars still exist after the maximum-mass point for negative values of $\beta$. However, the minimum binding energy is compatible with the zero frequency of the fundamental pulsation mode. }  
\end{figure*}

Let us now focus on $\omega_0^2$ since it is the lowest eigenvalue of all the allowed pulsation modes. For a range of central densities, in Fig.~\ref{SFFigMIT} we plot the squared frequency of the fundamental mode as a function of the central density (left panel) and gravitational mass (right plot) for quark matter. Similar to the pure GR case, $\omega_0^2$ is always a decreasing function of the central density for quark stars. For a fixed central density, a positive (negative) value of $\beta$ leads to an increase (decrease) in $\omega_0^2$. According to the right plot, it can be observed that the maximum mass does not correspond to $\omega_0^2= 0$ when $\beta \neq 0$. In other words, the existence of stable quark stars is possible after the maximum-mass configuration for negative values of $\beta$, while the onset of instability is indicated before reaching the maximum mass for positive values of $\beta$. Therefore, the classical $M(\rho_c)$ method widely used to analyze the stellar stability in GR is not compatible with the calculation of vibration mode frequencies in Rastall gravity. Nevertheless, we remark that the cusp formed by plotting the binding energy as a function of proper mass can be used as an indicator for the onset of radial instability of relativistic stars in Rastall gravity. Specifically, we observe that the central density where the binding energy is minimal corresponds exactly to $\rho_c$ where $\omega_0^2= 0$ for all considered values of $\beta$.

\begin{figure*}
 \includegraphics[width=8.56cm]{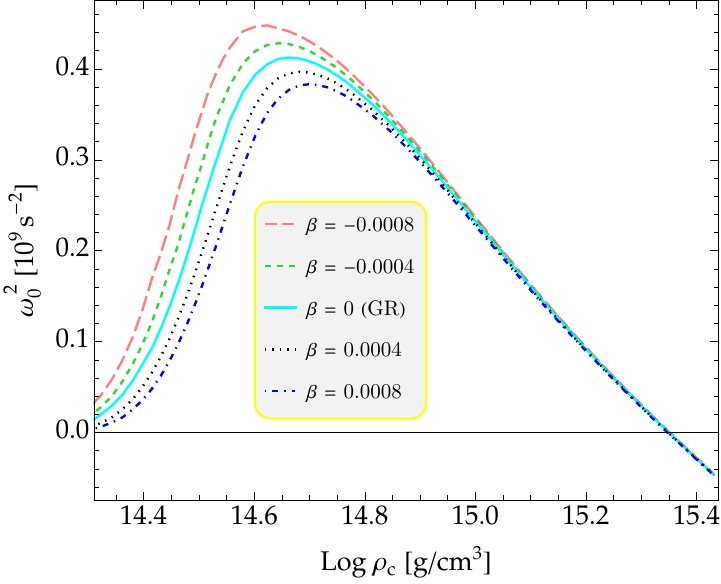}
 \includegraphics[width=8.628cm]{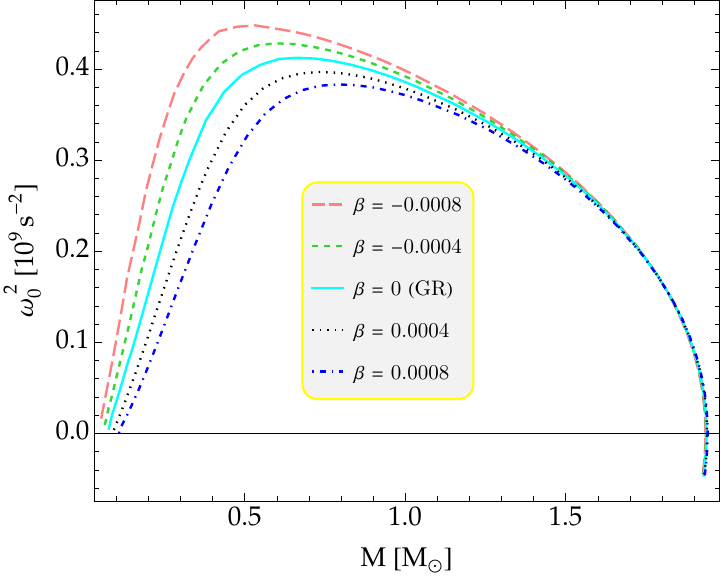}
 \caption{\label{SFFigIUFSU} Squared frequency of the fundamental pulsation mode against the central density (left panel) and gravitational mass (right panel) for hadronic matter in Rastall gravity. It can be observed that the largest impact of $\beta$ on the eigenvalue $\omega_0^2$ occurs in the low-central-density region, while it is practically unchanged in the high-mass region and, as a consequence, the maximum-mass value almost corresponds to $\omega_0^2= 0$. Notice that the behavior here is different from the case of quark stars as shown in Fig.~\ref{SFFigMIT}. }  
\end{figure*}

Unlike quark stars, the squared frequency of the fundamental mode for hadronic matter grows until it reaches a maximum value and then decreases with increasing central density, see Fig.~\ref{SFFigIUFSU}. The increase in parameter $\beta$ leads to a significant decrease in $\omega_0^2$ at sufficiently low central densities, while the changes are irrelevant in the high-central-density region. Notice that this behavior is contrary to the case of quark stars. Nonetheless, since the values of $\vert\beta\vert$ are smaller than those considered in the quark case, it is not possible to appreciate substantial variations in the maximum-mass values. In this direction, we could conclude that the maximum-mass configuration indicates (approximately) the transition between stable and unstable neutron stars in Rastall gravity.

\section{Conclusions}\label{Sect5}

Within the context of Rastall gravity, we have derived the modified TOV equations and we investigated the hydrostatic equilibrium of compact stars by adopting two equations of state. In the case of neutron stars, the radius undergoes a substantial modification due to the Rastall parameter $\beta$ in the low-mass region, namely, positive values of $\beta$ increase the radius and vice-versa. However, for quark stars it was necessary to use larger values of $\beta$ in order to observe appreciable changes in the mass-radius diagram. The radius of these stars increases (decreases) for negative (positive) values of $\beta$, contrary to the results obtained using the hadronic matter EoS. The maximum mass for quark stars can be significantly increased by considering negative values of $\beta$. Furthermore, the gravitational redshift of light emitted at the stellar surface exhibits considerable changes only in the high-mass region for quark stars.

Following a perturbative procedure, similar to that carried out by Chandrasekhar in Einstein's theory, we have derived for the first time the differential equations governing the adiabatic radial pulsations in Rastall gravity in order to examine the dynamical stability. In the GR limit (i.e., when $\beta \rightarrow 0$) we recover the corresponding Chandrasekhar equations. Under suitable boundary conditions, the system of equations has been treated as an eigenvalue problem, and vibration frequencies have been determined for a wide range of central densities for both quark and neutron stars. Our numerical results revealed that the standard $M(\rho_c)$ criterion for stellar stability is no longer valid in Rastall gravity because the maximum-mass point does not correspond to $\omega_0^2= 0$. Indeed, we found that there exist stable quark stars after the maximum-mass configuration for negative values of $\beta$.

As an independent approach, we have also analyzed the gravitational binding energy as a function of proper mass for a whole family of quark stars using several values of $\beta$. Our findings showed that the critical central density corresponding to the maximum-mass configuration deviates from $\rho_c$ corresponding to the minimum-binding energy. However, the concept of binding energy has been shown to be useful in constructing stable compact stars since the squared frequency of the fundamental oscillation mode vanishes at the central-density value corresponding to the minimum-binding-energy configuration. Therefore, the cusp formed when the binding energy is minimal can be used as a turning point from stability to instability of compact stars in Rastall gravity.

\begin{acknowledgments}
JMZP acknowledges support from ``Fundação Carlos Chagas Filho de Amparo à Pesquisa do Estado do Rio de Janeiro'' -- FAPERJ, Process SEI-260003/000308/2024.
\end{acknowledgments}\

\textbf{Data Availability Statement:} This work is of theoretical nature and the numerical data generated have been included in the manuscript through plots.

\end{document}